\documentclass[aps,prd,twocolumn,groupedaddress, showpacs]{revtex4} % 2 columns

\usepackage{graphicx}

\begin{document}

% \preprint{}

\title{Large curvature perturbations near horizon crossing \\
in single-field inflation models}



\author{Edgar Bugaev}
\email[e-mail: ]{bugaev@pcbai10.inr.ruhep.ru}
%\thanks{}

\author{Peter Klimai}
\email[e-mail: ]{pklimai@gmail.com}
%\thanks{}
\affiliation{Institute for Nuclear Research, Russian Academy of
Sciences, 60th October Anniversary Prospect 7a, 117312 Moscow,
Russia}


% !-! We do not put date to arXiv
%\date{\today}

\begin{abstract}
We consider the examples of single-field inflation models
predicting large amplitudes of the curvature perturbation power
spectrum at relatively small scales. It is shown that in models
with an inflationary potential of double-well type the peaks in
the power spectrum, having, in maximum, the amplitude ${\cal
P}_{\cal R} \sim 0.1$, can exist (if parameters of the potential
are chosen appropriately). It is shown also that the spectrum
amplitude of the same magnitude (at large $k$ values) is predicted
in the model with the running mass potential, if the positive
running, $n'$, exists and is about $0.005$ at cosmological scales. Estimates of the
quantum diffusion effects during inflation in models with the
running mass potential are given.
\end{abstract}

\pacs{98.80.Cq, 04.70.-s \hfill arXiv:0806.4541 [astro-ph]}
% 98.80.Cq = Particle-theory and field-theory models of the early Universe
% 04.70.-s = Physics of black holes

%\keywords{}

\maketitle

\section{Introduction}

In last few months several papers appeared \cite{Kohri:2007gq,
Kohri:2007qn, Saito:2008em, Peiris:2008be}, in which single-field
inflation models predicting (potentially) large amplitudes of the
curvature perturbations on relatively small scales are discussed.
It is shown in \cite{Kohri:2007gq} that large class of such models
exists, namely, the models with a potential of hill-top type (the
idea of the hill-top inflation was proposed, to author's
knowledge, in the earlier work \cite{Boubekeur:2005zm}). In such
models, the potential can be of concave-downward form at
cosmological scales (in accordance with data) and be much flatter
at the end of inflation when small scales leave horizon.
Correspondingly, the amplitude of the perturbation power spectrum
can be rather large. It is noticed in \cite{Kohri:2007gq} that the
running mass model, having the potential with the similar
behavior, also can predict the large spectrum amplitude.

Authors of \cite{Kohri:2007qn} discuss also more general scenarios
of producing large amplitudes of perturbation spectrum. They show
the limitedness of the standard procedure of the potential
reconstruction which can easily miss the potentials leading to
large spectrum amplitude and to noticeable primordial black hole
(PBH) production.

In the recent paper \cite{Saito:2008em} it was shown that PBH
production is possible in single-field models of two-stage type
("chaotic $+$ new"). The idea was proposed ten years ago in
\cite{Yokoyama:1998pt}. Authors of \cite{Saito:2008em} carried out
the numerical calculation of the power spectrum using the
Coleman-Weinberg (CW) potential.

In the present paper we continue a study of the problems discussed
in the previous works \cite{Kohri:2007gq, Kohri:2007qn,
Saito:2008em}. We investigated thoroughly, as a particular
example, the model of two-stage inflation with a potential of the
double-well (DW) form, and showed that the characteristic features
of the power spectrum in models of this type (such as an amplitude
and a position of the peak, a degree of tuning of parameters of
the potential) are very sensitive to an exact form of the
potential. Further, we carried out the numerical calculation of
the power spectrum in a running mass model and showed that the
spectrum amplitude at small scales can be rather large. Our
calculation differs from the previous one \cite{Leach:2000ea} in
several aspects: we express the results through the values of
parameters $s$, $c$, which are used nowadays and prove to be very
convenient for a comparison with data; we studied, in details, the
difference in predictions of slow-roll and numerical approaches at
high $k$-values; we exactly specified the value of the positive
running, $n'$, which corresponds to our spectrum prediction. In
the final part of the work we investigated the quantum diffusion
effects in a model with the running mass potential.

A plan of the paper is as follows. In the
second section we study predictions of two-stage inflation models
with DW and CW potentials, with accent on a mechanism of the
formation of peaks in the power spectrum. In the Sec. \ref{sec-RM}
all aspects connected with an obtaining of the predictions of
running mass inflation models are discussed. In the Sec.
\ref{sec-Concl} we present our main conclusions.

\begin{figure}
\includegraphics[width=0.9\columnwidth]{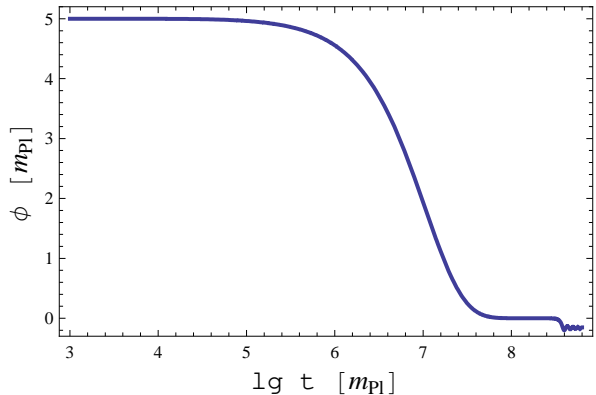} %
\center \caption{ \label{DWbkg} The solution of the background
equation for inflation with the double-well potential (\ref{DW}).
The parameters of the
potential are: $v = 0.16286748 m_{Pl}, \lambda=1.7\times 10^{-13}$.} %
\end{figure}
%%%%%%%%%%%%%%%%%%%%%%%%%%%%%%%%%%%%%%%%%%%%%%%%%%%%%%%%%%%%%
\begin{figure}[!t]
\includegraphics[width=0.9\columnwidth]{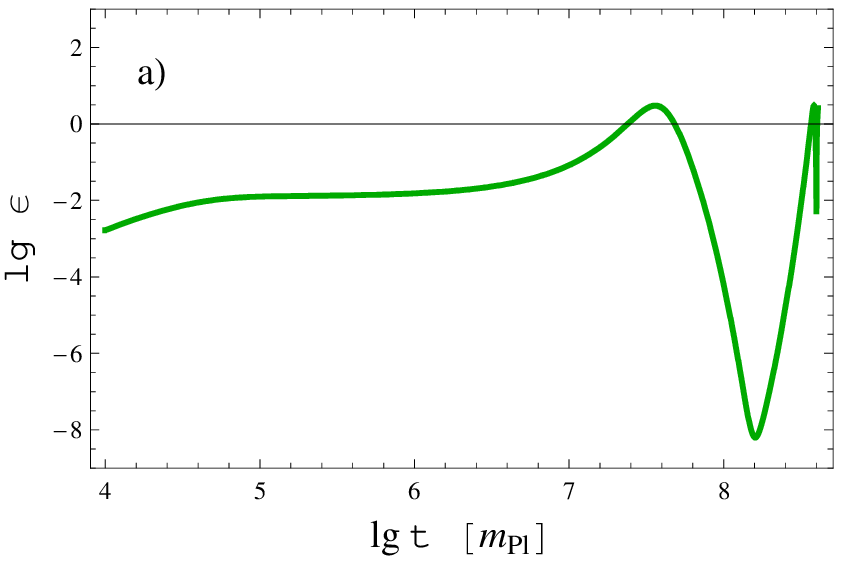}
\includegraphics[width=0.9\columnwidth]{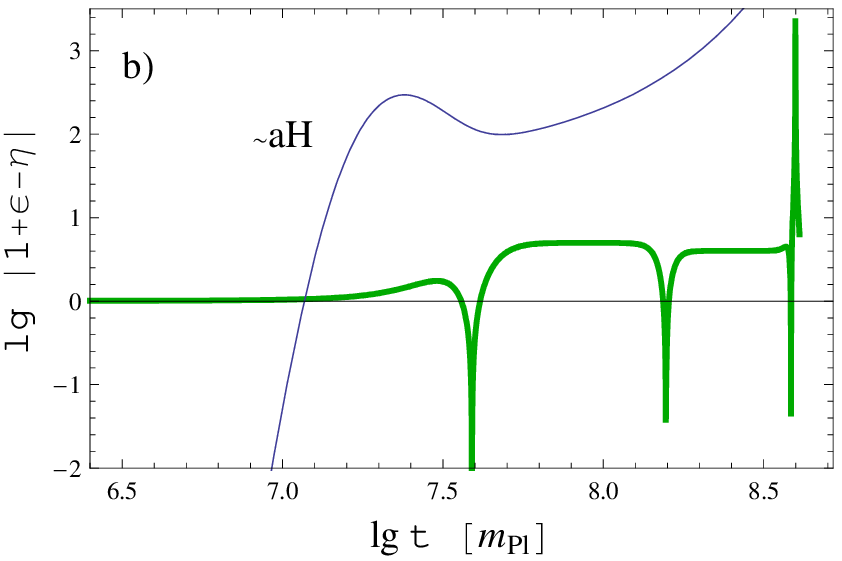}
\center %
\caption{ \label{DWed} The time dependence of the parameter
$\epsilon$ and the combination $1+\epsilon-\eta$ corresponding to
the background field evolution
shown in fig. \ref{DWbkg} } %
\end{figure}
%%%%%%%%%%%%%%%%%%%%%%%%%%%%%%%%%%%%%%%%%%%%%%%%%%%%%%%%%%%%%

\section{Examples of the power spectrum with peaks}

\subsection{Double-well potential}

This form of the inflaton potential having an unstable local
maximum at the origin has been discussed many times in studies of
eternal and new inflation. The main problem was to realize the
initial condition for the new inflation when system starts from a
top of the hill. Ten years ago the model of "chaotic new
inflation" has been proposed \cite{Yokoyama:1998pt}, in which the
system climbs on the top during dynamical evolution of the
inflaton field with initial conditions coinciding with those of
chaotic inflation models. In the approach of
\cite{Yokoyama:1998pt} the inflation has two stages, chaotic and
new, and during transition from the first stage to the second the
slow-roll conditions break down (in general).

The potential has two parameters:
\begin{equation}
V(\phi) = \frac{\lambda}{4} (\phi^2-v^2)^2 \;. \label{DW}
\end{equation}
The inflaton starts with the rather high value of $\phi$ (we take
$\phi_{\rm in} \sim 5 m_{Pl}$) and rolls down to the origin. The
parameter $\lambda$ is fixed by a normalization of the power
spectrum on experimental data, $\lambda \sim 10^{-13}$. The
evolution of the system strongly depends on the value of $v$: if
$v$ is finely tuned, $\phi$ can spend some time near the origin,
i.e. on the top, and then roll down to one of the two minima. In
figs. \ref{DWbkg} and \ref{DWed}a the time evolution for the
inflaton and the parameter $\epsilon$ for the definite values of
the parameters $\lambda$ , $v$ are shown. One can see that,
really, $\phi \approx 0$ at some period of time and, what is
important, the slow-roll approximation is invalid ($\epsilon \sim
1$) just at the time of the transition from a rolling to a
temporary stay at the top of the potential.

It is well known that in situations when there is a failure of the
slow-roll evolution the perturbations on super-horizon scales can
be amplified and specific features in the power spectrum can arise
\cite{Starobinsky:1992ts, Ivanov:1994pa, Bullock:1996at, Leach:2000yw, Leach:2001zf}
(see also the recent paper \cite{Jain:2007au}). In particular, in the earliest work
where this problem was studied \cite{Starobinsky:1992ts}, the inflation potential with a sudden
gradient discontinuity leading to the power spectrum of a step-like form was considered.
All this means that the
predictions of the slow-roll approximation which are based on the
assumption that perturbations reach an asymptotic regime outside
the horizon cannot be trusted.

The curvature perturbation on comoving hypersurfaces ${\cal R}_k$, as a function of the
conformal time $\tau$, is a solution of the differential equation
(prime denotes the derivative over $\tau$)
\begin{equation}
{\cal R}_k'' + 2 \frac{z'}{z} {\cal R}_k'  + k^2 {\cal R}_k = 0 ,
\label{Rkpp}
\end{equation}
\begin{equation}
\frac{z'}{z} = aH(1+\epsilon-\eta) \;\; , \;\; z \equiv \frac{a\dot\phi}{H} \;  \label{zpz}
\end{equation}
($\phi$ is the inflaton field). The standard initial condition for this equation,
corresponding to the Bunch-Davies \cite{Birrell:1982ix} vacuum, is
\begin{equation}
u_k(\tau) = \frac{1}{\sqrt{2 k}}\; e^{- i k \tau} \;\; , \;\; aH
\ll k \;,
\end{equation}
where $u=z {\cal R}$. The variable $u$ had been introduced in \cite{Lukash2, Mukhanov2, Sasaki:1986hm}.

The functions  $\epsilon$ and $\eta$ in eq.
(\ref{zpz}) are the Hubble slow-roll parameters defined by the expressions \cite{Liddle:1994dx}
\begin{equation}
\epsilon = - \frac{\dot H}{H^2} = \frac{4 \pi}{m_{Pl}^2}
 \frac {\dot \phi^2}{H^2} \;\;\; ,  \;\;\; %
\eta = - \frac {\ddot \phi } {H \dot \phi} \; .
\label{epsetaformula}
\end{equation}
Outside the slow-roll limit these functions are not necessarily small.

It had been demonstrated in \cite{Leach:2000yw} that solutions of
the equation (\ref{Rkpp}),
at $k \ll aH$, i.e., outside horizon,
are well approximated by constant if the coefficient of the
friction term, $z'/z$, doesn't change sign near the horizon
crossing. In the opposite case, if $z'/z$ changes sign at some
time, the friction term becomes a negative driving term, and one
can expect strong effects on modes which left horizon near that
time. In the present paper we study the corresponding features of
the power spectrum, following closely the analysis of
\cite{Leach:2000yw}.

According to eq. (\ref{zpz}), $z'/z$ is proportional to
$1+\epsilon-\eta$ and the comoving Hubble wave number $aH$. The
time dependences of these functions are shown in fig. \ref{DWed}b.
One can see that the interruption of inflation correlates with the
change of the sign of $1+\epsilon-\eta$.

\begin{figure}
\includegraphics[width=0.98\columnwidth]{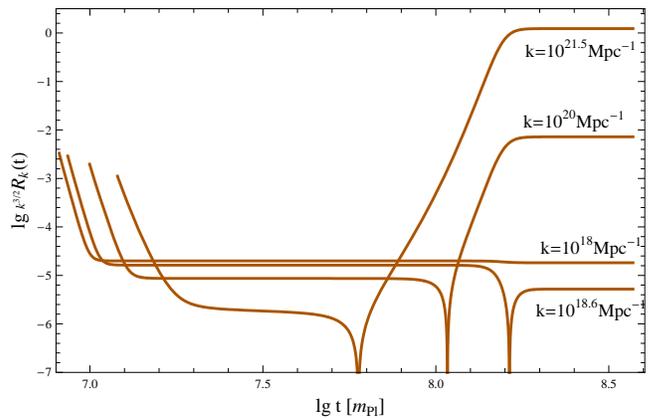} %
\center \caption{ \label{DW-Rk-diff-k} A time evolution of the
curvature perturbation ${\cal R}_k(t)$ for several different
values of wave number $k$ during inflation with the DW potential.
The parameters of the potential are the same as
in fig. \ref{DWbkg}.  } %
\end{figure}

\begin{figure}
\includegraphics[width=0.98\columnwidth]{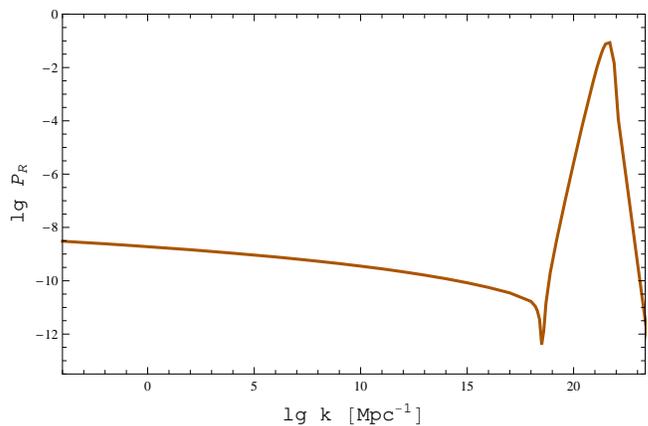} %
\center \caption{\label{DW-PR} The numerically calculated power
spectrum ${\cal P}_{\cal R}(k)$ for the model with the potential
(\ref{DW}). Parameters of the potential used in the calculation
are the same as in fig. \ref{DWbkg}.  } %
\end{figure}

The time evolution of curvature perturbations for several modes is
shown in fig.  \ref{DW-Rk-diff-k}. It is clearly seen that the
perturbations ${\cal R}_k$ for different modes freeze out at
different amplitudes. The mode which crosses horizon near the
moment of time when the sign of $1+\epsilon-\eta$ changes (i.e.,
near $t \approx 7.5 m_{Pl}$) freezes at maximum amplitude, due to
the exponentially growing driving term in eq. (\ref{Rkpp}) (which
is most effective just for this mode). It leads to the
characteristic peak in the power spectrum ${\cal P}_{\cal R}(k)$,
\begin{equation}
{\cal P}_{\cal R} (k) = \frac{4\pi k^3}{(2\pi)^3} | {\cal R}_k
|^2,
\end{equation}
shown in fig. \ref{DW-PR}.

The calculations of ${\cal R}_k$ (fig. \ref{DW-Rk-diff-k}) are
carried  out up to the end of inflation, and the power spectrum in
fig. \ref{DW-PR} also corresponds to this moment of time. We
estimate approximately the reheat temperature in our case as $\sim
(\lambda v^4)^{1/4} \sim 10^{14}$ GeV. The horizon mass at the
beginning of radiation era is
\begin{equation}
M_{hi} \sim 10^{17} {\rm g} \left( \frac{10^7 {\rm GeV} }{T_{\rm
RH} } \right) ^2 \sim 10^3 {\rm g} \;,
\end{equation}
and maximum wave number, which equals the Hubble radius at the end
of inflation, is
\begin{equation}
k_{\rm end} = a_{\rm eq} H_{\rm eq} \left( \frac{M_{\rm
eq}}{M_{hi}}  \right)^{1/2} \sim 10^{23} \; {\rm Mpc}^{-1} .
\end{equation}

\begin{figure}
\includegraphics[width=0.98\columnwidth]{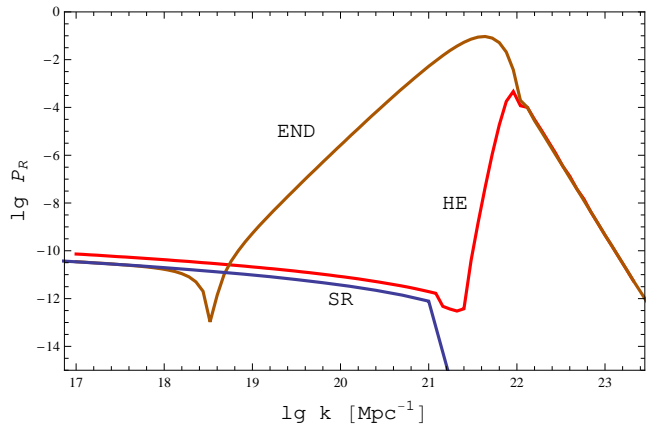} %
\center \caption{\label{DW-PR-diff} The power spectrum ${\cal
P}_{\cal R}(k)$ for the model with potential (\ref{DW}). END:
${\cal P}_{\cal R}(k)$ is calculated at the end of inflation; HE:
${\cal P}_{\cal R}(k)$ is calculated at
the time of the horizon exit; SR: the slow-roll result.} %
\end{figure}

In the fig. \ref{DW-PR-diff} we show the power spectrum calculated
for two stages  of its evolution: at horizon exit (HE) and at the
end of inflation (END). In the same figure the result of the
calculation with an use of the slow-roll formulae is also shown.
The peak of HE curve at the region of large $k$ is due to a
failure (for ${\cal R}_k$) to reach the asymptotic limit. One can
see also that the slow-roll approximation is too crude to describe
perturbations at the end of inflation (this our conclusion agrees
with general statements of \cite{Leach:2000yw}).

%%%%%%%%%%%%%%%%%%%%%%%%%%%%%%%%%%%%%%%%%%%%%%%%%%%%%%%%%%%%%%%%%%%%%%%%%%%%
%%%%%%%%%%%%%%%%%%%%%%%%%%%%%%%%%%%%%%%%%%%%%%%%%%%%%%%%%%%%%%%%%%%%%%%%%%%%
%%%%%%%%%%%%%%%%%%%%%%%%%%%%%%%%%%%%%%%%%%%%%%%%%%%%%%%%%%%%%%%%%%%%%%%%%%%%
%%%%%%%%%%%%%%%%%%%%%%%%%%%%%%%%%%%%%%%%%%%%%%%%%%%%%%%%%%%%%%%%%%%%%%%%%%%%

\subsection{Coleman-Weinberg potential}

The Coleman-Weinberg potential has the form \cite{Coleman:1973jx}:
\begin{equation}
V(\phi) = \frac{\lambda}{4}\phi^4 \left( \ln \Big | \frac{\phi}{v}
\Big|
 - \frac{1}{4} \right) +   %
\frac{\lambda}{16} v^4 .\label{CW}
\end{equation}
It looks very similar to the previous one, but the important
difference is its behavior near the origin. Namely, the CW
potential behaves as $A + B \phi^4 \ln (\phi / v)$ near the
origin, i.e., it is more flat near zero, in comparison with the DW
potential. Therefore, it has more e-folds of "new inflation"
\cite{Yokoyama:1998pt} and, as a consequence, the peaks of the
power spectrum (arising, as in the previous case, due to the
temporary interruption of inflation) correspond to relatively
smaller $k$ values. Besides, at the beginning of new inflation when the very flat region of the potential near
zero is crossed by $\phi$, quantum fluctuations with a particle creation can be large (e.g., see below,
Sec. \ref{qde}) and must be taken into account.

\begin{figure}
\includegraphics[width=0.98\columnwidth]{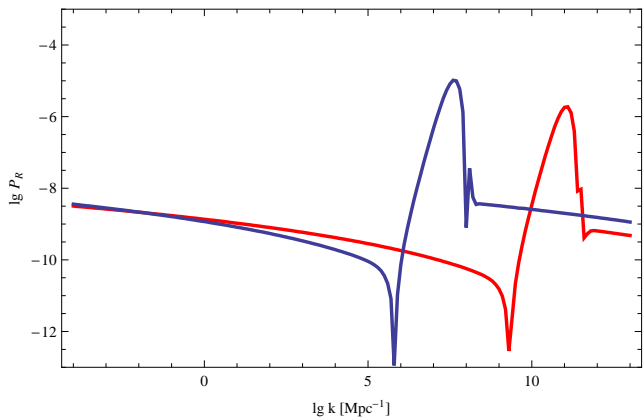}
\center %
\caption{\label{CW-PR} The result for the power spectrum
${\cal P}_{\cal R}(k)$ calculation for the CW potential
(\ref{CW}), for two sets of parameters.  Left peak is for $v =
1.113 M_P, \lambda = 5.5\times 10^{-13}$.  For the right peak, $v
= 1.112 M_P, \lambda = 2.4\times 10^{-13}$. }
\end{figure}

In fig. \ref{CW-PR} two examples of the power
spectrum calculations are shown for two different sets of
parameter values. As before, the peaks are very distinct, although
their amplitudes are smaller.

%%%%%%%%%%%%%%%%%%%%%%%%%%%%%%%%%%%%%%%%%%%%%%%%%%%%%%%%%%%%%%%%%%%%%%%%%%%%%%%%%%%%%%%%%%%%%%%
%%%%%%%%%%%%%%%%%%%%%%%%%%%%%%%%%%%%%%%%%%%%%%%%%%%%%%%%%%%%%%%%%%%%%%%%%%%%%%%%%%%%%%%%%%%%%%%

\subsection {Possibilities of PBH production}

One can see, in particular, from fig. \ref{DW-PR}, that, in
principle, the production of primordial black holes (about these
objects, see original works \cite{Zeldovich1967, Hawking:1971ei}
and reviews \cite{Carr:2005zd, Khlopov:2008qy}) can be rather
large in single-field inflation models with potentials of
double-well type. The main conclusion is that it requires rather
large fine tuning of parameters of the potential. The
characteristic PBH mass is estimated by
\begin{equation}
M_{\rm BH} \approx M_h = M_{hi} \left( \frac{k_{\rm end}}{k_{\rm
peak}}  \right) ^ 2 \; ,
\end{equation}
and in the case of the spectrum of fig. \ref{DW-PR}, $M_{\rm BH}
\sim 10^7$g. In the CW  case, fig. \ref{CW-PR}, $M_{\rm BH} \sim
100 M_{\odot}$ for the left peak and $M_{\rm BH} \sim 10^{27}$g
for the right peak (but the amplitudes of the spectra are too
small).

Recently, it has been shown \cite{Saito:2008em} that inflation
with CW  potential is capable to produce significant number of
PBHs: the parameter $v$ can be chosen (by finest tuning!) in such
a way that inflaton field makes several oscillations from one
minimum to another before it climbs on the top and "new inflation"
starts.

In the present paper we do not consider possibilities of a
constraining of the peak amplitudes, based, for example, on the
effects of PBH evaporation in early Universe (see, e.g.,
\cite{Bugaev:2006fe}).

\section{ Running mass model \label{sec-RM} }

\subsection { Main assumptions and approximations}

We consider in more detail a case of the running mass inflation
model \cite{Stewart:1996ey, Stewart:1997wg, Covi:1998mb,
Covi:1998jp, Covi:1998yr, German:1999gi, Covi:2004tp} which
predicts a spectral index with rather strong scale dependence. The
potential in this case takes into account quantum corrections in
the context of softly broken global supersymmetry and is given by
the formula
\begin{equation}
V = V_0 + \frac{1}{2} m^2(\ln \phi) \phi^2.
\end{equation}
The dependence of the inflaton mass on the renormalization scale
$\phi$ is  determined by the solution of the renormalization group
equation (RGE).

{\bf 1.} The inflationary potential in supergravity theory is of
the order of $M_{\rm inf}^4$, where $M_{\rm inf}$ is the scale of
supersymmetry breaking during inflation. In turn, the mass-squared
of the inflaton (and any other scalar field) in supergravity has,
in general, the order of the square of Hubble expansion rate
during inflation,
\begin{equation}
| m^2 | \sim H_I^2 = \frac{V_0}{3 M_P^2}.
\end{equation}
We suppose, for simplicity (see \cite{Stewart:1996ey,
Stewart:1997wg, Covi:1998jp, Covi:1998yr}), that $M_{\rm inf} \sim
M_{\rm s}$, where $M_{\rm s}$ is the scale of supersymmetry
breaking in the vacuum,
\begin{equation} M_{\rm s} \sim \sqrt{\tilde m_s M_P} \sim
10^{11} {\rm GeV} \sim 3\times 10^{-8}  M_P
\end{equation}
($\tilde m_s$ is the scale of squarks and slepton masses, $\tilde
m_s\sim 3$ TeV). These assumptions give the scale of the
inflationary potential:
\begin{equation}
V_0 \sim M_{\rm s}^4 \sim 10^{-30} M_P^4 \;\;\; , \;\;\; H_I
\approx 10^{-15} M_P.
\end{equation}

{\bf 2.} RGE for the inflaton mass is the following (we consider a
model \cite{Covi:1998jp, Covi:1998yr} of hybrid inflation using
the softly broken SUSY with gauge group $SU(N)$ and small Yukawa
coupling):
\begin{equation}
m^2(t) = m_0^2 - A \tilde m_0^2 \left[ 1 -
\frac{1}{(1+\tilde\alpha_0 t)^2} \right] \;\;\; , \;\;\; t \equiv
\ln \frac{\phi}{M_P} \; , \label{m2}
\end{equation}
$m_0^2$ and $\tilde m_0^2$ are, correspondingly, the inflaton and
gaugino  masses at $\phi=M_P$,
\begin{equation}
\tilde\alpha_0 = \frac{B \alpha_0}{2 \pi} ,
\end{equation}
$\alpha_0$ is the gauge coupling constant, $\alpha_0=g^2/4 \pi$.
$A$ and $B$  are positive numbers of order $1$, which are
different for different variants of the model, even if they are
based on the same supersymmetric gauge group $SU(N)$ (it depends
on a form of the superpotential, particle content of
supermultiplets, etc). We use in the present parer the variant of
\cite{Covi:1998yr} and, correspondingly, put everywhere below
$A=2$ and $B=N=2$.

{\bf 3.} A truncated Taylor expansion of the potential around the
particular scale  $\phi_0$ (in our case, $\phi_0$ is the inflaton
value at the epoch of horizon exit for the pivot scale $k_0
\approx 0.002h$ Mpc$^{-1}$) is
\begin{equation}
V(\phi) = V_0 + \frac{\phi^2}{2} \left[ m^2( \ln (\phi_0) ) - c
\frac{V_0}{M_P^2} \ln \frac{\phi}{\phi_0} + ... \right].
\end{equation}
Here, constant $c$ is defined by the equation
\begin{equation}
c \frac{V_0}{M_P^2}= \left. - \frac{dm^2}{d \ln \phi}
\right|_{\phi = \phi_0} . \label{ccc1}
\end{equation}

In turn, a Taylor expansion of eq. (\ref{m2}) up to linear terms
gives ($t_0 = \ln \frac{\phi_0}{M_P}$):
\begin{equation}
m^2(t) = m^2(t_0) - 4 \tilde m_0^2 \frac{ \tilde \alpha_0}
{(1+\tilde \alpha_0 t_0)^3 } \ln \frac{\phi}{\phi_0} \; .
\label{m2m2}
\end{equation}
From eqs. (\ref{ccc1}) and (\ref{m2m2}) we obtain the expression
for  the constant $c$,
\begin{equation}
c \frac{V_0}{M_P^2}= 4 \tilde m_0^2 \frac {\tilde \alpha_0}
{(1+\tilde \alpha_0 t_0)^3 } . \label{ccc2}
\end{equation}

If $|m_0^2| \sim \tilde m_0^2 \approx H_I^2$, then
\begin{equation}
c=\frac{4}{3} \frac{\tilde \alpha_0} {(1+\tilde \alpha_0 t_0)^3 }
.
\end{equation}

%%%%%%%%%%%%%%%%%%%%%%%%%%%%%%%%%%%%%%%%%%%%%%%%%%%%%%%%%%%%%%%%%%%%%%%%%%%%%%%%%%%%%%%%%%%%%%%
\begin{figure}
\includegraphics[width=0.7\columnwidth]{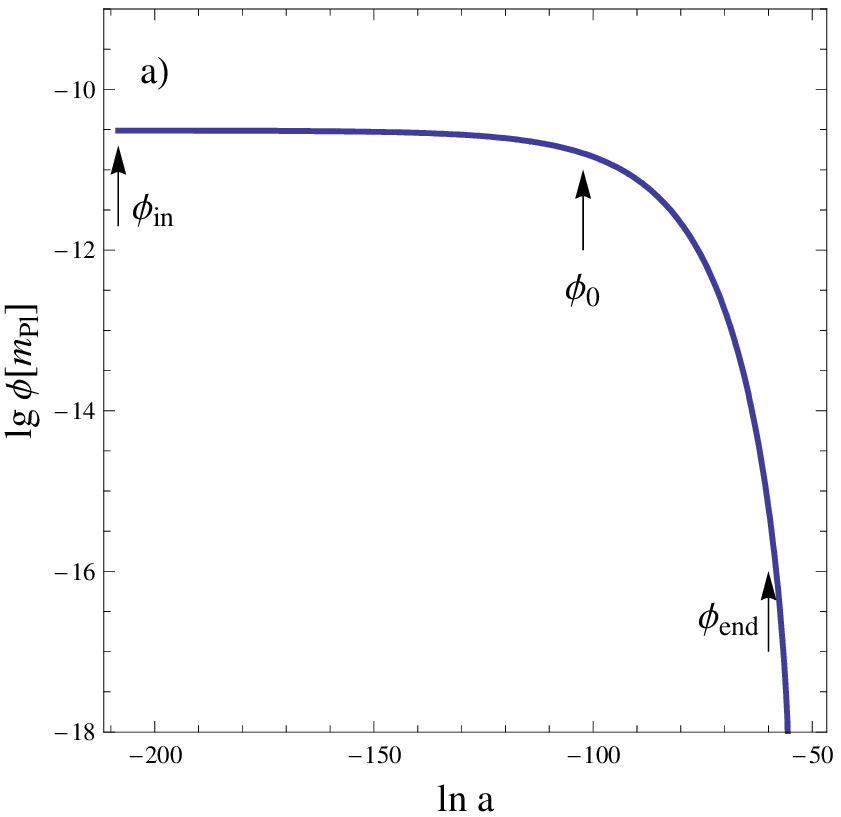}
\includegraphics[width=0.7\columnwidth]{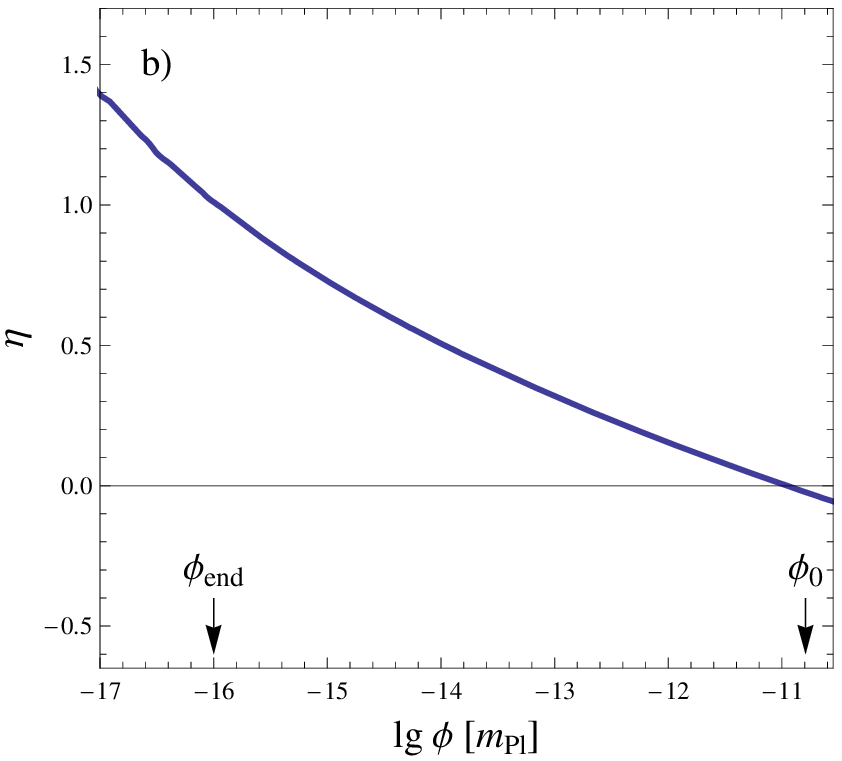}%
\center %
\caption{ \label{RM-phi-a-eta} {\bf a)} Evolution of  the
inflaton field $\phi(\ln a)$ in the running mass model. {\bf b)}
The dependence of the parameter $\eta$ on a value of the field
$\phi$. For both plots,
$H_I=10^{-15} M_{P}$, $c = 0.062$, $s = 0.040$. } %
\end{figure}
%%%%%%%%%%%%%%%%%%%%%%%%%%%%%%%%%%%%%%%%%%%%%%%%%%%%%%%%%%%%%%%%%%%%%%%%%%%%%%%%%%%%%%%%%%%%%%%

It appears (see fig. \ref{RM-phi-a-eta}b) that in our example
$\phi_0 \sim 10^{-10} M_P$, so, $t_0 \sim \ln 10^{-10} \sim
(-23)$. Assuming that $\alpha_0 \sim 1/24$ (as in SUSY-GUT
models), one has $\tilde \alpha_0 \sim
\frac{2}{2\pi}\frac{1}{24}$. In such a case, $c \sim 4 \tilde
\alpha_0 \sim 0.06$.

If we would keep terms of higher order in $t-t_0=\ln
\frac{\phi}{\phi_0}$ in the Taylor expansion of $m^2(t)$ in eq.
(\ref{m2m2}) we would see that the real expansion parameter is
$\tilde \alpha_0 \ln \frac{\phi}{\phi_0}$ rather than $\ln
\frac{\phi}{\phi_0}$. The smallest value of $\phi$, $\phi_{\rm
end}$, in our case is $\sim 10^{-16}M_P$ (see fig.
\ref{RM-phi-a-eta}b). Even for such value of $\phi_{\rm end}$, the
expansion parameter is rather small,
\begin{equation}
\tilde \alpha_0 \ln \frac{\phi_{\rm end}}{\phi_0} \sim \tilde
\alpha_0 \ln 10^{-6} \sim (-0.1) \; .
\end{equation}
Having this in mind, we will use the linear approximation for the
inflaton  mass (eq. (\ref{m2m2})) in the entire region of inflaton
field values exploited in the present paper.

Following the previous papers, we introduce also another
parameter,
\begin{equation}
s = c \ln \left( \frac{\phi_*}{\phi_0} \right),
\end{equation}
where $\phi_*$ is the inflaton value corresponding to a maximum of
the potential. This parameter connects the field value $\phi_0$
with the Hubble parameter during inflation and with the
normalization of the CMB power spectrum:
\begin{equation}
\phi_0 s = \frac{H_I}{2\pi {\cal P}_{\cal R}^{1/2} (k_0) } .
\label{phi0PR}
\end{equation}

{\bf 4.} The minimum value of the inflaton field which corresponds
to the end of inflation can be determined from the approximate
equation \cite{Covi:1998yr}
\begin{equation}
\eta = M_P^2 \frac{V''}{V} \cong \frac{M_P^2}{V_0} m^2 =1 \;.
\end{equation}
Using RGE, one obtains from this formula the relation
\begin{equation}
\frac{M_P^2}{V_0} \left( m_0^2 - A \tilde m_0^2 +  \frac{A \tilde
m_0^2}{(1+\tilde \alpha_0 t)^2 } \right) = 1.
\end{equation}
Substituting here $A=2$, $\tilde m_0^2 = |m_0^2| = V_0/3M_P^2$,
one has finally the approximate expression for $\phi_{\rm end}$,
\begin{equation}
\phi_{\rm end} = M_P  \exp\left[{-\frac{1}{\tilde\alpha_0} \left(
1 - \frac{1}{\sqrt{3}} \right) }\right] \;, \label{phial0}
\end{equation}
which shows that the minimum field value is very sensitive to the
value of the model parameter $\tilde\alpha_0$ and, in our case,
does not depend on $V_0$. More exactly, the condition $\eta = 1$
means  the end of the {\it slow-roll part} of inflation. We
suppose, as usual (see, e.g. \cite{Stewart:1996ey,
Stewart:1997wg}) that in reality inflation ends by hybrid
mechanism, and the critical value of inflaton field, $\phi_{\rm
cr}$, is determined by the value of the Yukawa coupling $\lambda$
(in spite of the inequality $\lambda^2 \ll \alpha$). One can check
\cite{Covi:1998yr} that the value of $\lambda$ can always be
chosen such that $\phi_{\rm cr} < \phi_{\rm end}$ and slow-roll
ends before the reaching of $\phi_{\rm cr}$.

One should note that the accuracy of the approximate formula
(\ref{phial0}) is not  very good. Luckily, in the approach based
on the numerical calculation of the power spectrum there is no
need to use it, because the value of $\phi_{\rm end}$ appears in a
course of the calculation (fig. \ref{RM-phi-a-eta}b).

\subsection {Power spectrum of curvature perturbations }

An analysis of CMB anisotropy data \cite{Spergel:2006hy,
Dunkley:2008ie}, including other types of observation
\cite{Lesgourgues:2007te}, leads to the following main qualitative
conclusions:

{\it i)} the power spectrum of scalar curvature perturbations is
red, i.e., the spectral index is negative,
\begin{equation}
n_0 = 0.97 \pm 0.01 \; ;
\end{equation}

{\it ii)} observations are consistent, or, at least, are not in
contradiction with the small positive running of the spectral
index, $n_0' < 0.01$;

{\it iii) } the contribution of tensor perturbations in the value
of the spectral index is small ($\lesssim 10^{-2}$) and, as a
result, $n \approx 1+2\eta$ ; it means that $\eta$ is negative,
and the potential must be concave-downward (i.e., of hill-top
type), while cosmological scales cross horizon during inflation
\cite{Kohri:2007gq}.

One should note that, strictly speaking, the conclusion {\it iii)} is not grounded firmly enough.
According to the recent analysis \cite{Lesgourgues:2007aa}, the present data still admit any sign of $\eta$ and $V''$.

These conclusions constrain the possible values of the parameters
$s$ and $c$. Approximately, for cosmological scale one has
\begin{equation}
n_0 - 1 \approx 2 (s - c) \;\; , \;\; n_0' \approx 2 s c \;.
\end{equation}
From the conclusion {\it iii)} it follows that $c>0$ (it is
consistent with eq. (\ref{phi0PR})), from the positivity of $n_0'$
(the conclusion {\it ii)} ) it follows that $s>0$. At last, the
conclusion {\it i)} leads to the inequality $s<c$.

We choose for the power spectrum calculation the following values:
\begin{equation}
c= 0.062 \;\; , \;\; s=0.040 .
\end{equation}
These numbers correspond, at cosmological scales, to the following
values of slow-roll parameters:
\begin{equation}
\epsilon \approx \frac{s \phi_0^2} {M_P^2} \sim 10^{-21} \;\; ;
\;\; \eta \approx s - c \sim (-0.02) \;,
\end{equation}
that seems to be consistent with the present data
\cite{Peiris:2008be}.

To check the validity of the slow-roll approximation, we calculate
the spectrum by the three ways: {\it i)} using the approximate
analytic slow-roll formula
\begin{eqnarray} \label{PRaprf}
\frac{{\cal P}_{\cal R} (k)}{{\cal P}_{\cal R} (k_0) } = \exp{
\left[ \frac{2 s}{c} \left( e^{c \Delta N(k) } -1  \right) - 2 c
\Delta N(k) \right] }
\end{eqnarray}
($\Delta N(k)\equiv \ln(k/k_0)$; this expression is easily derived
from the simplest slow-roll prediction
\begin{equation}
{\cal P}_{\cal R} (k) = \left. \frac{H^2}{\pi \epsilon m_{Pl}^2}
\right| _ {aH=k} \; ,  \label{SRformula}
\end{equation}
which gives the power spectrum to leading order in the slow-roll
approximation \cite{Liddle:1993fq}); {\it ii)} using the
Stewart-Lyth approximation \cite{Stewart:1993bc}, which is valid
to first order in the slow-roll approximation,
\begin{eqnarray} \label{SLformula}
{\cal P}_{\cal R}^{1/2} (k) = \left[1-(2C+1)\epsilon + C \eta
\right] \frac{1}{2 \pi} \left. \frac{H^2}{| \dot\phi|} \;
\right|_{aH=k} \; , \\  C \approx - 0.73 \; ; \nonumber
\;\;\;\;\;\;\;\;\;\;\;\;\;\;\;\;\;\;\;\;\;\;\;\;\;\;
\end{eqnarray}
{\it iii)} by numerical integration of the differential equation for
${\cal R}_k$, eq. (\ref{Rkpp}).

%%%%%%%%%%%%%%%%%%%%%%%%%%%%%%%%%%%%%%%%%%%%%%%%%%%%%%%%%%%%%%%%%%%%%%%%%%%%%%%%%%%%
\begin{figure}
\includegraphics[width=0.94\columnwidth]{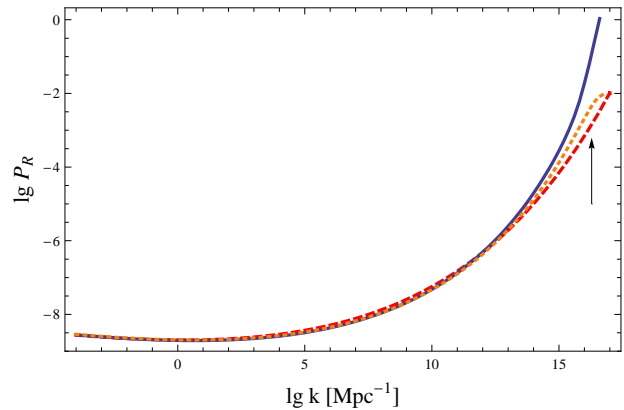} %
\center %
\caption{ \label{PR-RM} Power spectrum ${\cal P}_{\cal R} (k)$ in
the running mass model, calculated numerically (solid line), by
the approximate analytic formula (\ref{PRaprf}) (long-dashed line)
and using the Stewart-Lyth extended slow-roll approximation
(short-dashed line). The parameters of the potential are the same
as used in fig. \ref{RM-phi-a-eta}. The arrow shows the value of
$k_{\rm end}$. }
\end{figure}
%%%%%%%%%%%%%%%%%%%%%%%%%%%%%%%%%%%%%%%%%%%%%%%%%%%%%%%%%%%%%%%%%%%%%%%%%%%%%%%%%%%%%

\begin{figure}
\includegraphics[width=0.9\columnwidth]{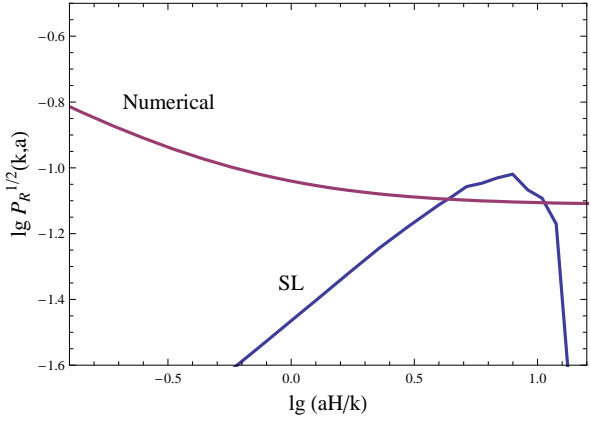} %
\center \caption{\label{SLnum} Dependence of ${\cal P}_{\cal R}
(k, a)$ calculated numerically and ${\cal P}_{\cal R}$ from the
Stewart-Lyth formula. The comoving wave number for this figure is
$k=10^{15.8}$ Mpc$^{-1}$. }
\end{figure}

The results of the calculations are presented in figs.
\ref{RM-phi-a-eta}-\ref{SLnum}. Fig. \ref{RM-phi-a-eta} shows the
evolution of the inflaton field $\phi$ with the scale factor and a
growth of the slow-roll parameter $\eta$ with a decrease of $\phi$
from $\phi_0$ to $\phi_{\rm end}$. The power spectrum is shown in
fig. \ref{PR-RM} for a broad interval of comoving wave numbers. It
is clearly seen that near the end of inflation, when
\begin{equation}
\phi \sim 10^{-16} M_P \;\; , \;\; k\sim k_{\rm end} =  a_{\rm
end}H_{\rm end} = 3 \times 10^{16} {\rm Mpc}^{-1} \;,
\end{equation}
the slow-roll formulae are inaccurate: they strongly underestimate
values of ${\cal P}_{\cal R}$. To illustrate this point more
clearly, at the next figure we show the comparison of two curves:
$aH$-dependence of the spectrum calculated numerically for a
definite value of $k$, $k=10^{15.8}$ Mpc$^{-1}$, and $aH$-
dependence of the Stewart-Lyth spectrum. It is seen that the
numerical spectrum at the moment of crossing horizon (when $aH=k$)
is already almost asymptotical, and its value distinctly exceeds
the corresponding value predicted by Stewart-Lyth formula.

%%%%%%%%%%%%%%%%%%%%%%%%%%%%%%%%%%%%%%%%%%%%%%%%%%%%%%%%%%%%%%%%%%%%%%%%%%%%%%%%%%%%%
%%%%%%%%%%%%%%%%%%%%%%%%%%%%%%%%%%%%%%%%%%%%%%%%%%%%%%%%%%%%%%%%%%%%%%%%%%%%%%%%%%%%%

\subsection {Quantum diffusion effects}
\label{qde}

We calculated the curvature perturbations in terms of the
classical trajectories of a scalar field associating, in
particular, points in a field space with definite numbers of
e-folds from the end of inflation. This description becomes
incorrect if the quantum diffusion destroys the classical
evolution of the field. In this case we should use the methods of
stochastic inflation. The latter approach operates with the
coarse-grained field, which is defined to be spatial average of
the field  $\phi$ over a physical volume with size larger than the
Hubble radius $H^{-1}$.

In the slow-roll approximation, the evolution of the
coarse-grained field  $\varphi$ is governed by the first order
Langevin-like equation \cite{Starobinsky:1982ee, Starobinsky:1986fx, Vilenkin2,
Linde:1986fd}

\begin{equation}
\dot \varphi + \frac{1}{3H}V'(\varphi) = \frac{H^{3/2}}{2\pi}
\xi(t), \label{Lang}
\end{equation}
\begin{equation}
\langle \xi(t) \rangle = 0 \; , \; \langle \xi(t) \xi(t')\rangle =
\delta(t-t').
\end{equation}
Here, $\xi(t)$ is a random noise field, and angular brackets mean
ensemble average. The term $\frac{1}{3H}V'(\varphi)$ describes the
deterministic evolution of the field $\varphi$, in the absence of
the noise term $\frac{H^{3/2}}{2\pi} \xi(t)$. The solution of eq.
(\ref{Lang}), in the absence of the noise term, is the
deterministic slow-roll trajectory $\varphi_{\rm sr}(t)$. Going to
finite time differences, the coefficient $\frac{H^{3/2}}{2\pi}$
can be rewritten as $\sqrt{\frac{H^3}{4 \pi^2 \Delta t} }$, and
the evolution of $\varphi$ on timescales $\Delta t \ge H^{-1}$ can
be described by a finite-difference form of the eq. (\ref{Lang}),
\begin{equation}
\varphi(t+\Delta t) - \varphi(t) = - \frac{1}{3H}V'(\varphi)
\Delta t +   \frac{1}{2 \pi} \sqrt{H^3 \Delta t} \; \xi(t).
\label{LangFD}
\end{equation}
The condition for the deterministic evolution is (see, e.g.,
\cite{Winitzki:2006rn})
\begin{equation}
\frac{1}{3H} | V'(\varphi) | \Delta t \gg \frac{1}{2 \pi}
\sqrt{H^3 \Delta t} \;\; , \;\; \Delta t = H^{-1}.
\end{equation}
Using the slow-roll connection between $V$ and $H$, one obtains
\begin{equation}
| V'(\varphi) | \gg \frac{3}{2 \pi} H^3 = \frac{1}{2 \pi \sqrt{3}}
V^{3/2}(\varphi). \label{compareV}
\end{equation}

In the approach of \cite{Martin:2005ir} the coarse-grained field
is considered as a perturbation of the classical solution
$\varphi_{\rm cl}$ (which is the solution of the Langevin equation
without the noise),
\begin{equation}
\label{phi012} \varphi(t) = \varphi_{\rm cl}(t) + \delta
\varphi_{1}(t) +  \delta \varphi_{2}(t) + ... \; .
\end{equation}
Here, the term $\delta \varphi_{i}(t)$ depends on the noise at the
power $i$.  It is assumed that the Hubble parameter in the
Langevin equation depends only on the coarse-grained field
$\varphi$,
\begin{equation}
H^2(\varphi) = \frac{1}{3 M_P^2} V(\varphi).
\end{equation}
Correspondingly, the Hubble parameter can be expanded
perturbatively,
\begin{equation}
H(\varphi) = H_{\rm cl} + H_{\rm cl}'(\delta \varphi_{1} + \delta
\varphi_{2})  + \frac{H_{\rm cl}''}{2} \delta \varphi_{1}^2 + ...
\; ,
\end{equation}
\begin{equation}
H_{\rm cl} = H(\varphi_{\rm cl}) =  \sqrt {\frac{V(\varphi_{\rm
cl})}{3 M_P^2}
 }.
\end{equation}

%%%%%%%%%%%%%%%%%%%%%%%%%%%%%%%%%%%%%%%%%%%%%%%%%%%%%%%%%%%%%%%%%%%%%%%%%%%%%%%%%%%%
\begin{figure} [!t]
\includegraphics[width=0.9\columnwidth]{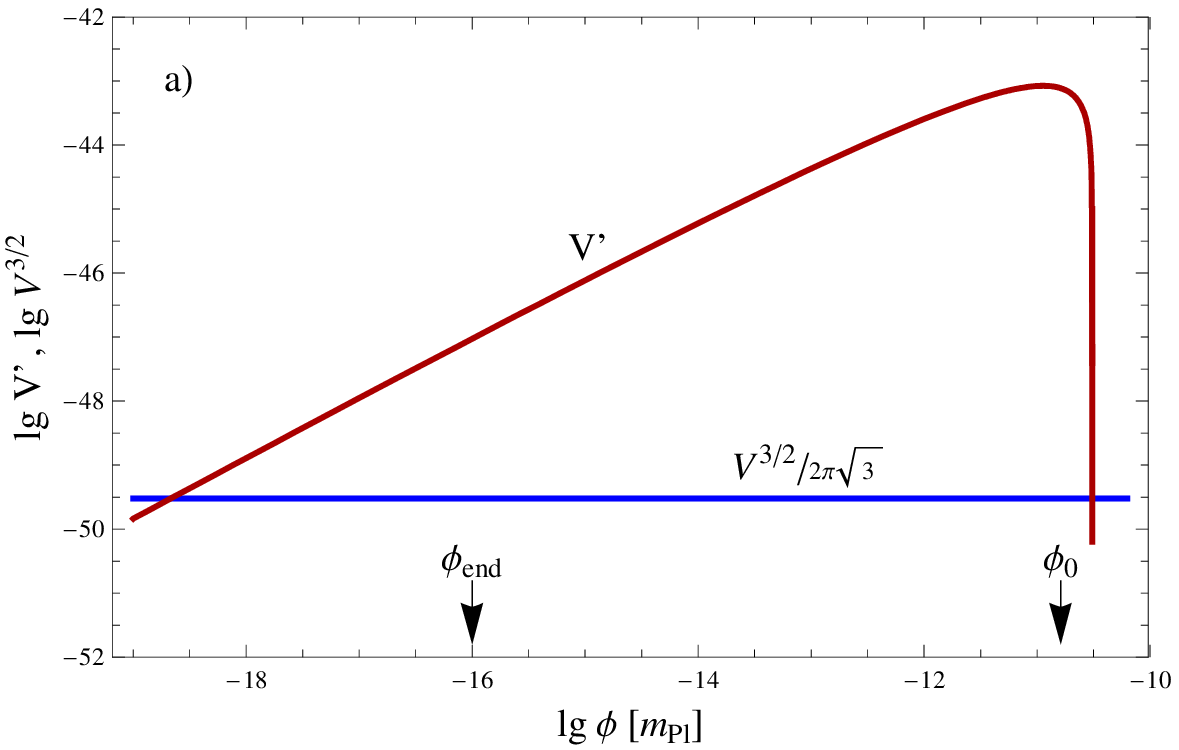}
\includegraphics[width=0.9\columnwidth]{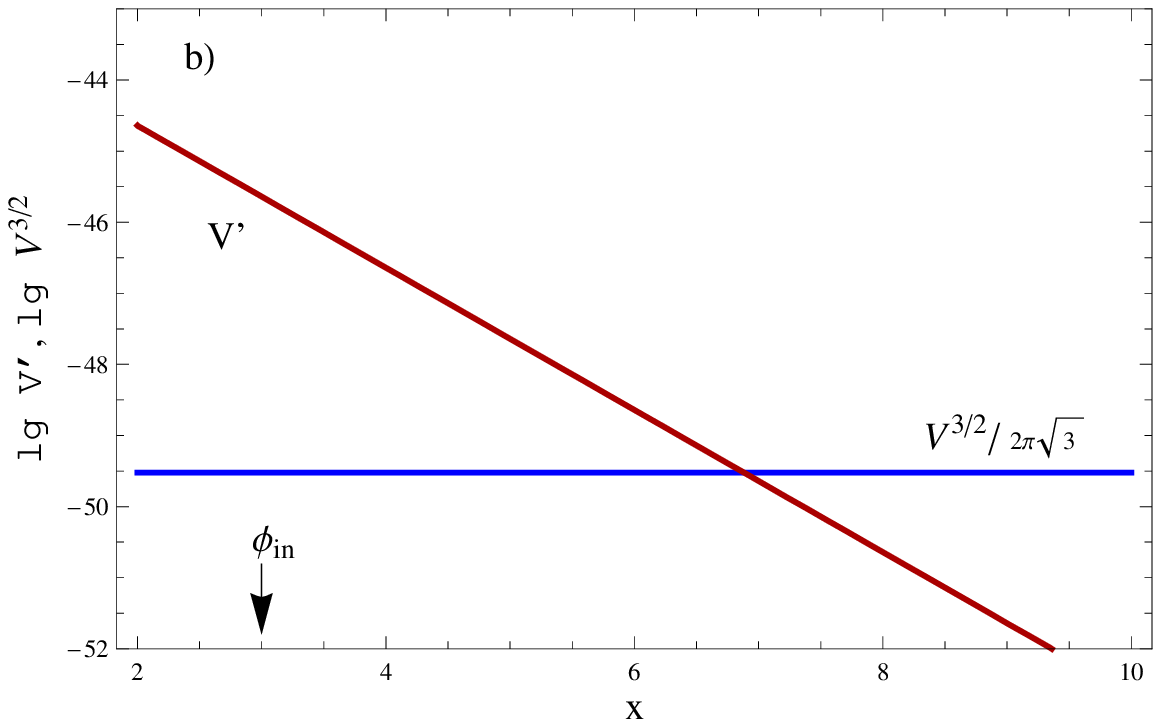} %
\center \caption{ \label{compV-fig} {\bf a)} Comparison of $V'$
and $V^{3/2}$ for the running mass model (see eq.
(\ref{compareV})). {\bf b)} The same figure in different scale:
variable $x$ is connected to the field value by the relation
$\varphi= (1- 10^{-x})\varphi_*$. } %
\end{figure}

%%%%%%%%%%%%%%%%%%%%%%%%%%%%%%%%%%%%%%%%%%%%%%%%%%%%%%%%%%%%%%%%%%%%%%%%%%%%%%%%%%%%
\begin{figure} [!t]
\includegraphics[width=0.98\columnwidth]{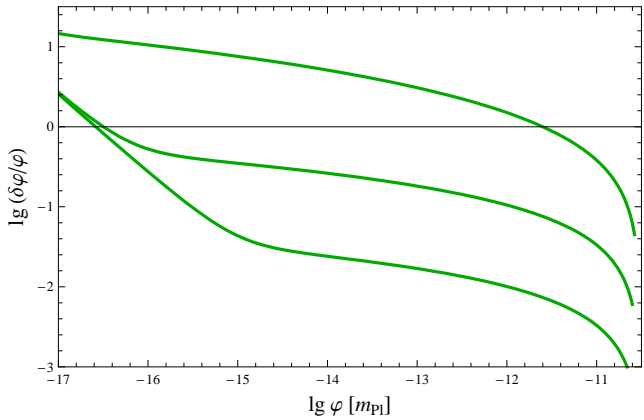}  %
\center \caption{ \label{dff-fig} The result for the calculation
of $\delta \varphi / \varphi$ for different values of
$\varphi_{\rm in} = \varphi_*(1-10^{-x})$. From bottom to top, $x$
equals 3, 4, 5. Here, $\delta \varphi \equiv \sqrt {\langle
\delta\varphi_1^2 \rangle} +\langle \delta\varphi_2 \rangle$.
} %
\end{figure}
%%%%%%%%%%%%%%%%%%%%%%%%%%%%%%%%%%%%%%%%%%%%%%%%%%%%%%%%%%%%%%%%%%%%%%%%%%%%%%%%%%%%

This approach permits to calculate the mean value of the total
number  of e-folds, $\langle N \rangle$, and to compare it with
the corresponding "classical" number,
\begin{eqnarray} \nonumber
N_T^{\rm cl} = - \frac{1}{2 M_P^2} \int\limits_{\varphi_{\rm
in}}^{\varphi_{\rm end}}  d \varphi_{\rm cl} \frac{H_{\rm
cl}}{H_{\rm cl}'} =
\;\;\;\;\;\;\;\;\;\;\;\;\;\;\;\;\;\;\;\;\;\;\;\;\;\;\;
\\
= \frac{1}{M_P^2} \int\limits_{\varphi_{\rm
end}}^{\varphi_{\rm in}}  d \varphi_{\rm cl} \left(  \frac{V}{V'}
\right)  ;
\end{eqnarray}
\begin{eqnarray} \nonumber
\delta N_T = \langle N_T \rangle - N_T^{\rm cl} =
\;\;\;\;\;\;\;\;\;\;\;\;\;\;\;\;\;\;\;\;\;\;\;\;\;\;\;\;\;\;\;\;\;\;\;\;
\\
= - \frac{1}{2
M_P^2}  \int\limits_{\varphi_{\rm in}}^{\varphi_{\rm end}} \left[
\langle \delta \varphi_2 \rangle + \frac{H_{\rm cl}''}{2 H_{\rm
cl}'} \langle \delta \varphi_1^2 \rangle \right] d \varphi_{\rm
cl}. \label{NT}
\end{eqnarray}
Besides, one can calculate the mean value of the Gaussian
probability  distribution function for the coarse-grained field
and to see how it behaves as a function of the current field
value.

The interval of inflaton field values for which the inequality
(\ref{compareV}) holds and, therefore, the deterministic evolution
dominates, is shown in fig. \ref{compV-fig}a,b. The variable $x$
in fig. \ref{compV-fig}b is defined by the relation
\begin{equation}
\frac{\varphi_{\rm in}}{\varphi_{*}} = 1 - 10^{-x}
\label{x-define}
\end{equation}
($\varphi_{*}$ is, as before, the point of a maximum of the
potential $V(\varphi)$). It is seen from the figure that the
constraints on the inflaton field following from the condition
(\ref{compareV}) are not too severe,
\begin{equation}
3 \times 10^{-19} M_P \;\; \lesssim \;\;  \varphi \;\;  \lesssim
\;\; \varphi_* - 10^{-7} \varphi_* \;. \label{phiinlimit}
\end{equation}

An accuracy of the perturbative expansion (\ref{phi012}) for the
monomial potential of stochastic inflation models was studied in
\cite{Martin:2005hb}. Here we study this accuracy for the running
mass potential, using the parameters $V_0$, $s$, $c$,
$\varphi_{\rm end}$ introduced above. The results of the
calculation of $\langle \delta\varphi_1^2 \rangle$ and $\langle
\delta\varphi_2 \rangle$ are shown in fig. \ref{dff-fig}. As one
can see, the perturbative expansion is good if the starting value
of the inflaton field, $\varphi_{\rm in}$, is chosen to be not too
close to the value of $\varphi$ at the maximum of the potential.
More exactly, the parameter $x$, defined in eq. (\ref{x-define}),
must be smaller than $4 \div 4.5$. The starting value
$\varphi_{\rm in}$ corresponds to a beginning of the evolution,
i.e., $\delta\varphi_1(t_{\rm in}) = \delta\varphi_2(t_{\rm in}) =
0$.

If the evolution is deterministic, the correction to a classical
e-fold number  $N_T^{\rm cl}$ , given by eq. (\ref{NT}), is small.
To estimate analytically the upper limit for this correction, we
used the analytic expressions for $\langle \delta\varphi_1^2
\rangle$ and $\langle \delta\varphi_2 \rangle$ derived in
\cite{Martin:2005ir}, keeping in them leading terms only.
According to these expressions, the following inequalities hold:
\begin{eqnarray} \nonumber
\langle \delta\varphi_2 \rangle \;\; < \;\; %
\left( \frac{V_0}{M_P} \right)^4 \left( \frac{M_P}{\varphi_{\rm
in}} \right)^2 \frac{1}{\ln^2 \frac {\varphi_{\rm
in}}{\varphi_{*}} } \frac{\varphi}{M_P} \; \lesssim \\ \lesssim
10^{-10+2x} \varphi \;\; ,
\end{eqnarray}
\begin{equation}
\langle \delta\varphi_1^2 \rangle \;\; < \;\; %
\left( \frac{V_0}{M_P} \right)^4 %
\frac{1}{\ln^2 \frac {\varphi_{\rm in}}{\varphi_{*}} } \; \sim \;
10^{-30+2x} M_P^2 \; .
\end{equation}
Using these upper limits, one can estimate the corresponding upper
limits of two integrals in the expression for $\delta N_T$
(\ref{NT}). The result is the following:
\begin{eqnarray}
\frac{1}{M_P^2} \int \limits_{\varphi_{\rm end}}^{\varphi_{\rm
in}} \langle \delta\varphi_2 \rangle d \varphi_{\rm cl} \;\; <
\;\; 10^{-30+2x} \; ;
\\
\frac{1}{M_P^2} \int \limits_{\varphi_{\rm end}}^{\varphi_{\rm
in}} \frac{ \langle \delta\varphi_1^2 \rangle H_{\rm cl}''} {
H_{\rm cl}' } d \varphi_{\rm cl} \;\; < \;\; 10^{-30+3x} \; .
\label{dNx}
\end{eqnarray}

It is clear from the inequalities (\ref{dNx}) that the quantum
correction to e-fold number, $\delta N_T$, is quantitatively small
even if the value of $x$ is as large as $10$. But only if $x <
4\div 4.5$ , and the perturbative expansion, eq. (\ref{phi012}),
is valid, one really can be sure that
\begin{equation}
\delta N_T \ll N_T^{\rm cl} \; ,
\end{equation}
and the evolution is deterministic.

%%%%%%%%%%%%%%%%%%%%%%%%%%%%%%%%%%%%%%%%%%%%%%%%%%%%%%%%%%%%%%%%%%%%%%%%%%%%%%%%%%%%
\begin{figure}[!t]
\includegraphics[width=0.9\columnwidth]{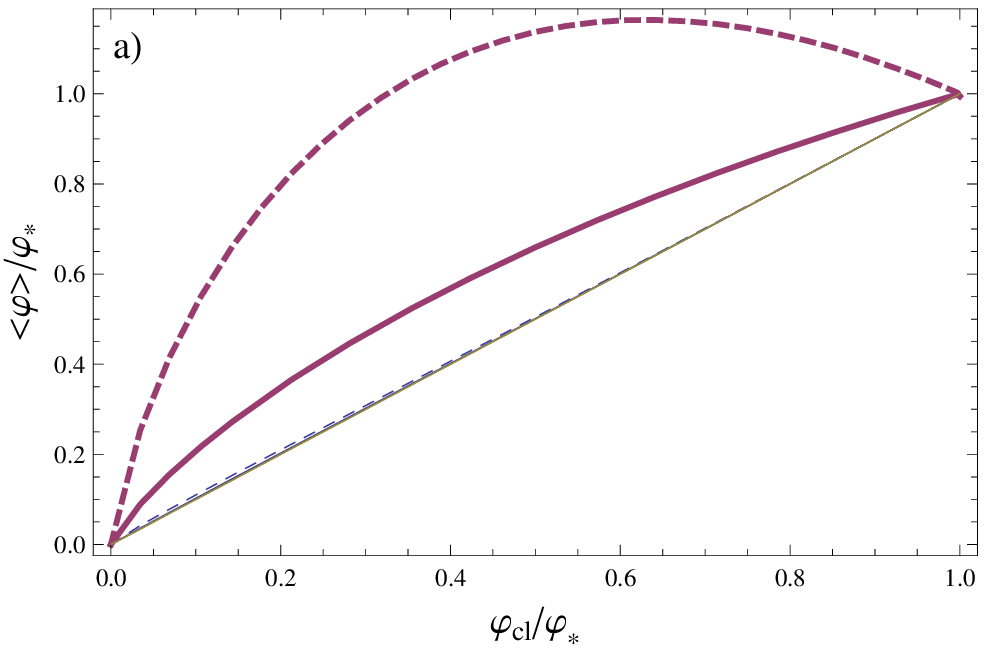}
\includegraphics[width=0.9\columnwidth]{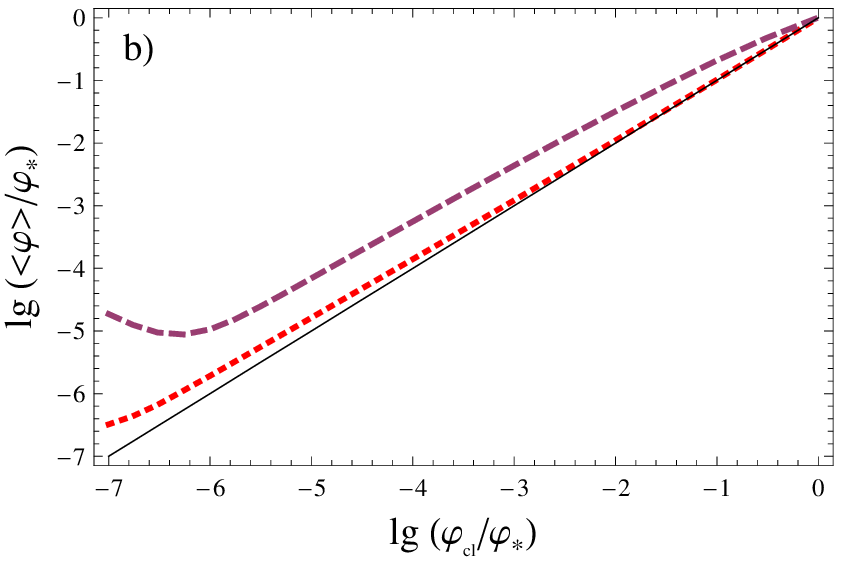} %
\center %
\caption{\label{fig-vol} Calculation of stochastic effects for
various values of initial
field $\varphi_{\rm in}$. %
{\bf a)} Thin dashed curve: $\varphi_{\rm in} = (1-1.5 \times
10^{-5})\varphi_*$, no volume effects included; thick dashed
curve: $\varphi_{\rm in}= (1-1.5 \times 10^{-5})\varphi_*$, volume
effects included; solid thick curve: $\varphi_{\rm in}=(1- 3
\times 10^{-5}) \varphi_*$ , volume effects included.
{\bf b)} Thick long-dashed and thick short-dashed curves
correspond to the cases with and without inclusion of volume
effects, respectively, for  $\varphi_{\rm in} = (1-3 \times
10^{-5}) \varphi_*$.
} %
\end{figure}
%%%%%%%%%%%%%%%%%%%%%%%%%%%%%%%%%%%%%%%%%%%%%%%%%%%%%%%%%%%%%%%%%%%%%%%%%%%%%%%%%%%%

However, this  analysis is still not complete: one must check also
the position of the mean value of the probability distribution
function for the coarse-grained field \cite{Martin:2005ir}. The
calculation, with taking into account the volume effects, leads to
the results shown in fig. \ref{fig-vol}a,b. It is seen from the
figure that in this case also, as in the calculation of the e-fold
number correction, the correct choice of the initial condition
plays a decisive role: there are no effects of a "walk" of the
mean value around the maximum of the potential (such effects were
noticed in \cite{Martin:2005ir}) if evolution starts from the
point which is far enough from the maximum ($x \lesssim 4.5$). Of
course, the realization of the initial condition of this kind is,
in itself, a problem. Supposedly, it could be provided by the
previous history of eternal inflation \cite{Stewart:1996ey,
Stewart:1997wg}.

\section{Conclusions \label{sec-Concl}}

{\bf 1.} It is shown, by numerical methods, that in the
single-field inflationary model with a simple double-well
potential the parameter values of this potential can be chosen in
such a way that the power spectrum of curvature perturbations
${\cal P}_{\cal R} (k)$ has a huge peak (with amplitude $\sim
0.1$) at large $k$ (and the right normalization and monotonic
behavior at cosmological scales). The peak arises due to temporary
interruption of the slow-roll near points of the minimum of the
potential, $\pm v$. The corresponding mass of PBHs produced in
early universe in a case of the realization of such power spectrum
is about $10^7$g.

The analogous behavior of the power spectrum was obtained by
authors of \cite{Saito:2008em} in a model with CW potential. There
are some important differences in the results of
\cite{Saito:2008em} and ours, in the peak amplitude and PBH mass,
connected, in particular, with a large flatness near the origin in
a case of the CW potential.

{\bf 2.} It is shown that the inflation model with running mass
potential predicts a rather large amplitude of the power spectrum
of curvature perturbations ($\sim 0.1$) at $k$-values $\sim
10^{16}$ Mpc$^{-1}$. For such a prediction, a very small positive
spectral index running at cosmological scales is necessary, $n'~
\sim 0.005$, as well as a small negative value for the slow-roll
parameter $\eta$ ($\approx -0.02$). Both this numbers do not
contradict with data. It is shown also that for an obtaining the
correct quantitative results for the power spectrum at largest
$k$-values an use of numerical methods is required because, in
general, slow-roll formulas are not accurate enough at the end of
inflation, where $\eta \approx 1$.

{\bf 3.} Quantum diffusion effects in a model with the running
mass potential are studied in details. It is shown that
inflationary evolution of the universe in a model with a scalar
field and the running mass potential can be described by the
classic deterministic equations, and for a possibility of such a
description the correct choice of the initial conditions is
crucial. Concretely, an initial value of the inflaton field (at
the beginning of the evolution) should not be too close to a point
of the maximum of the potential. If this condition is satisfied,
the quantum corrections to a total e-fold number and to a position
of the mean value of the probability distribution function are
small.

\begin{acknowledgments}
Authors are grateful to Prof. A.A. Starobinsky for useful remarks.

The work was supported by Russian Foundation for Basic Research
(grant 06-02-16135).
\end{acknowledgments}

%\bibliography{ms}

%%%%%%%%%%%%%%%%%%%%%%%%%%%%%%%%%%%%%%%%%%%%%%%%%%%%%%%%%%%%%%%%%%%%%%%%%%

\end{document}